\documentstyle[preprint,aps]{revtex}

\tighten

\begin{document}
\title{Proton decay of high-lying states in odd nuclei}

\author{Ch. Stoyanov}
\address{Institute for Nuclear Research and Nuclear Energy,\\*
boul. Tzarigradsko Chaussee 72,
1784 Sofia, Bulgaria}

\author{V. V. Voronov}
\address{Bogoliubov Laboratory of Theoretical Physics,\\*
Joint Institute for Nuclear Research,\\*
141980 Dubna, Moscow Region, Russia}

\author{N. Van Giai}
\address{Groupe de Physique Th\'eorique, Institut de Physique
Nucl\'eaire,\\*
F-91406 Orsay Cedex, France}

\vspace{1cm}
\date{November 1999}

\maketitle

\begin{abstract}
In the framework of the quasiparticle-phonon model, we study the
non-statistical proton decay of excited states in odd nuclei towards
low-lying collective states.
Partial cross sections and branching ratios for the proton decay of
the high angular momentum states in $^{41}$Sc, $^{59}$Cu and $^{91}$Nb
 are evaluated.
The calculated branching ratios predict strong direct proton decays to the
low-lying vibrational states in $^{41}$Sc and $^{91}$Nb.
A general agreement with existing experimental data is found.
\end{abstract}


\section{Introduction}

Nucleon  transfer reactions induced by hadronic probes at
intermediate energies favour the excitation of high angular
momentum states lying above particle emission threshold
\cite{G_S_V_88,For_90,Bea_94,For_95_1,For_95_2}.
The observed structures originate from the coupling
of the initial single-particle mode with more complex states.
The coupling of single-particle states with surface vibrations is
mainly responsible for the damping of the single-particle mode
\cite{G_S_V_88,G_S_91,Sol_92}.
The particle decay of highly excited states gives the opportunity
to study in detail the damping process. For example, the relative
contributions of the direct and statistical components to the damping
of single-particle mode can be found. Up to now, experimental
data and theoretical calculations are available mainly for the neutron decay of
high-lying single-particle modes. Very recently, the proton decays
of high-lying states in $^{41}$Sc, $^{59}$Cu and $^{91}$Nb have been measured
\cite{Yoo_98,Molen}.


In odd nuclei, the simplest excited states can be described as
admixtures
of single-particle states or quasiparticle states coupled to collective
excitations (or phonons) of the even-even core. This weak coupling
picture has been successfully applied to obtain the strength functions
of a variety of odd nuclei \cite{G_S_V_88,G_S_91}.
A method for calculating particle
escape widths in the framework of the quasiparticle-phonon model
(QPM)\cite{Sol_92,V_V_S_S_85} has been
suggested in Ref. \cite{G_S_91} where the inclusive semi-direct
neutron decays of high-lying states in $^{209}$Pb
have been studied  using the general procedure of Refs.
\cite{Y_A_86,G_S_90,Colo_92} which was established for the case of
nucleon emission from giant resonances. In a previous work
\cite{G_S_V_F_96}
we have
extended
the method of Ref.\cite{G_S_91} in order to calculate
non-statistical particle decays of
excited states in odd nuclei leading to exclusive channels which
correspond to the ground and
low-lying excited states of the even-even core.
In the present work we apply our method to study the proton decay.
Using the QPM we calculate the partial cross sections and branching
ratios for the proton decay of the high angular momentum states in
$^{41}$Sc, $^{59}$Cu and $^{91}$Nb
 and compare them with experimental data.

This paper is organized as follows: in Sec.\ II
we describe briefly our theoretical approach to treat the direct nucleon
decay of high angular momentum states of single-particle type.
In Sec.\ III a comparison of the calculated and measured
branching ratios for non-statistical proton decay of high-lying states
in $^{41}$Sc, $^{59}$Cu and $^{91}$Nb is presented.
Finally, in Sec.\ IV conclusions are drawn.

\section{ Theory }

\subsection {The projection operator method}

The projection operator method is a convenient approach to treat
problems involving single-particle continua in a many-body context.
The general formalism was introduced by Yoshida
and Adachi\cite{Y_A_86} and it has been applied to studies of
high-lying states in odd nuclei\cite{G_S_91,G_S_V_F_96}.
The detailed expressions can be found in Ref.\cite{G_S_V_F_96}.
Here, we simply recall the main features.

We write the hamiltonian of the A+1 system in the form:

\begin{eqnarray}
\noindent
H & = & h + H_{core} + H_{coupl}~.
\label{f1}
\end{eqnarray}

\noindent
The first term
describes the motion of a particle in an average potential $U$ created
by the particles in the core:
\begin{eqnarray}
h & = & -\frac{1}{2m}\nabla^2 + U~.
\label{f2a}
\end{eqnarray}
The core hamiltonian is a sum of single-particle hamiltonians
$h_i$ and two-body residual
interactions $V_{i,j}$:
\begin{eqnarray}
H_{core} & = & \sum_{i=1}^A h_i +\sum_{i<j}^A V_{i,j}~.
\label{f2b}
\end{eqnarray}
The last term of $H$ is a sum of interactions
between the odd particle and the core particles:
\begin{eqnarray}
H_{coupl} & = & \sum_{i=1}^A V_{0,i}~.
\label{f2c}
\end{eqnarray}

The physical spectrum of $h$ consists of a small number of
 bound states $\{\varphi_i, e_i \}$ and a continuum of
 scattering states $\{\varphi_e, e \}$ which form
altogether a complete, orthogonal basis. The projection
operator method consists in introducing another complete set of orthogonal
basis states which is a direct sum of two complementary subsets, a first subset of
discrete, orthonormal states $\{\phi_\alpha, \epsilon_\alpha \}$
which span the single-particle $q$ space and a second subset of
$\{\phi_\epsilon, \epsilon \}$ continuum states spanning the complementary
$p \equiv 1-q$ space.
We will denote by $a^\dagger_\alpha, a_\alpha$ ($a^\dagger_\epsilon, a_\epsilon$)
the creation and annihilation operators of state $\varphi_\alpha$
($\varphi_\epsilon$).

In the spirit of the QPM\cite{Sol_92} the
hamiltonian $H_{core}$ is treated in the random-phase approximation
(RPA) in a discrete space, i.e., the particle-hole configurations
of RPA are built only with q-space states. We denote by $E_\nu$ and
$O^\dagger_\nu$ the energies and creation operators of these RPA
states which describe core excitations. If $\mid 0 >$ represents the RPA
ground state of the core, the properties of the (A+1)-nucleus
except for its nucleon decay properties can be described in terms of
the one-particle states $a^\dagger_\alpha \mid 0 >$ and
one-particle-plus-phonon states $[a^\dagger_\beta\otimes
O^\dagger_\nu] \mid 0 >$. We can write:

\begin{eqnarray}
\mid d_i> & \equiv & d^\dagger_i \mid 0 >\nonumber\\
 & = & \Big(\sum_\alpha C^{(i)}_\alpha
a^\dagger_\alpha + \sum_{\beta,\nu} D^{(i)}_{\beta,\nu}
[a^\dagger_\beta\otimes O^\dagger_\nu ]\Big) \mid 0 >~.
\label{f5}
\end{eqnarray}

\noindent
We call Q space
the space spanned by the (real) state vectors $\mid d_i >$ and $Q$ the
corresponding projection operator.
The amplitudes $C^{(i)}_\alpha$ and
$D^{(i)}_{\beta,\nu}$~, and the energies $\omega_i$ of $\mid d_i >$
are determined by diagonalizing $H$ in the RPA, i.e., one solves:

\begin{eqnarray}
[QHQ, d^\dagger_i ] = \omega_i d^\dagger_i~,
\label{f6}
\end{eqnarray}

\noindent
within the approximation of commutator linearization. The distribution of
$\mid C^{(i)}\mid^2$ represents the strength function from which one
can deduce the spectroscopic factors\cite{G_S_V_88}.

To allow for nucleons to decay it
is necessary to introduce state vectors where the odd particle has a
non-zero probability of being at infinity. This is achieved by
constructing the P space complementary to Q space and consisting of
all states which are linear combinations of the following one-particle
and one-particle-plus-phonon configurations:

\begin{eqnarray}
\mid \epsilon > \equiv a^\dagger_\epsilon \mid 0 >~,~~ \mid \epsilon,
\nu > \equiv [a^\dagger_\epsilon \otimes O^\dagger_\nu ] \mid 0 >~.
\label{f7}
\end{eqnarray}

The present definition of P space neglects continuum
effects on the phonons $O^\dagger_\nu$ which can also in principle
couple to the continuum and emit nucleons by themselves. Actually,
these effects should have a small influence on the non-statistical
particle decay of the (A+1)-nucleus since the most important phonons
contributing to the particle-phonon coupling are the low-lying
collective states of the core.

The direct sum of Q space and P space is by construction the complete
particle-plus-phonon space in which the hamiltonian $H$ should be
solved. It is completely equivalent to solve in the simpler Q space
the more complicated effective hamiltonian:

\begin{eqnarray}
{\cal H} (E) & \equiv  & QHQ + QHP \frac {1}{E^{(+)}-PHP} PHQ
\nonumber\\
   & \equiv & H_{QQ} + W(E)~,
\label{f8}
\end{eqnarray}

\noindent
where $P$ is the projection operator onto P space ($P+Q=1$) and $E$ is
the energy of the system. The hamiltonian $\cal H$ is complex and
energy dependent. For each value of $E$ one has to find the set of
complex states and eigenenergies:

\begin{eqnarray}
\mid {\cal D}_i \rangle & \equiv & {\cal D}_i^\dagger \mid 0 \rangle~,
\nonumber\\
  &  & \nonumber\\
 {\Omega}_i & \equiv & {\bar \omega}_i - \frac{i}{2}\Gamma^{\uparrow}_i~,
\label{f9}
\end{eqnarray}

\noindent
satisfying:

\begin{eqnarray}
[{\cal H}(E) , {\cal D}^\dagger_i ] = \Omega_i {\cal D}^\dagger_i~.
\label{f10}
\end{eqnarray}

\subsection { Escape widths}
We consider a direct transfer reaction $a + A \to b+ (A+1)^*$
followed by a sequential decay $(A+1)^* \to p + A^*$
where the (A+1)-nucleus in a highly excited state
decays by a semi-direct proton emission. In this
process, the $\mid {\cal D}_i>$ states will
act as doorway states. If we describe the reaction mechanism in a
simple approach, e.g., a distorted wave Born approximation (DWBA), we
can write down the scattering amplitude from an initial channel $i$ where
the target $A$ is in its ground state $\mid 0>$ to a final channel $f$
where the residual nucleus is left in some excited state $\mid \nu> =
O^\dagger_\nu \mid 0>$ with excitation energy $E_\nu$ while the
escaping proton has an energy $E-E_\nu$. Using the complex
bi-orthogonal basis $\{ \mid {\cal D}_i>,<\bar {\cal D}_i\mid \}$, we have:

\begin{eqnarray}
T_{fi} & = & \sum_{lj}\sum_i \frac
{<\phi^{(-)}_{lj}(E-E_\nu),\nu \mid H \mid {\cal D}_i> <\bar
{\cal D}_i, b \mid V \mid 0, a>}{E-\bar \omega_i+ i\Gamma^{\uparrow}_i/2}~,
\label{w2}
\end{eqnarray}

\noindent
where $V$ is the interaction inducing the particle transfer from $a$
to $A$, $\phi^{(-)}$ is an incoming wave of $p$ space at energy
$E-E_\nu$, and the sum over $(l,j)$ is restricted by the angular
momenta and parities of states $\mid {\cal D}_i>$ and $\nu$.
The case where the final channel is the
ground state of the residual nucleus corresponds to the above
expression with $\nu=0, E_\nu=0$.

The nucleon-transfer matrix element $<\bar
{\cal D}_i, b \mid V \mid 0, a>$ is proportional to the one-quasiparticle
amplitude of the state $\mid {\cal D}_i>$. In analogy with the amplitude
$C^{(i)}$ of Eq.(\ref{f5}), we denote it by ${\cal C}^{(i)}$. We also
introduce the partial escape amplitudes of state $\mid {\cal D}_i>$ to
channel $\nu$:

\begin{eqnarray}
\gamma_{i,\nu}(lj) & \equiv & \sqrt{2\pi}
<\phi^{(-)}_{lj}(E-E_\nu),\nu \mid H \mid {\cal D}_i>~.
\label{w3}
\end{eqnarray}

The partial widths are:
\begin{eqnarray}
\Gamma_{i,\nu}^{\uparrow}\,\equiv \,\sum_{lj} \mid
\gamma_{i,\nu}(lj)\mid^{2}~.
\label{w4}
\end{eqnarray}

To obtain an expression
for the cross section simple enough to lend itself to a discussion in
terms of escape widths, let us assume furthermore that interference
terms between different doorway states can be neglected. The density
of $\vert {\cal D}_i \rangle$ states is large, and for each interval
centered around $E$ and containing $N$ states we define locally
averaged quantities:
\begin{eqnarray}
C^2(E) & \equiv & \sum_{i \in I} \vert C^{(i)}\vert^2/N~,\nonumber\\
\Gamma^{\uparrow}_\nu(E) & \equiv & \sum_{i \in I}
\Gamma^{\uparrow}_{i,\nu}/N~,\nonumber\\
\Gamma^{\uparrow}(E) & \equiv & \sum_{\nu} \Gamma^{\uparrow}_\nu(E)~.
\label{neweq}\end{eqnarray}
Then, one can rewrite the cross sections
in the following form:
\begin{eqnarray}
\sigma_{\nu}(E)\,\propto\,C^2(E)
\Gamma^{\uparrow}_\nu(E)/\Gamma^{\uparrow}(E)~.
\label{w5}
\end{eqnarray}
The branching ratios can be calculated by the following formula:
\begin{eqnarray}
B_{\nu}\,=\,\frac{\sigma_{\nu}(E)}{\sum_{\mu}\sigma_{\mu}(E)}\,=\,\frac{
\Gamma_\nu^{\uparrow}(E)}{\Gamma^{\uparrow}(E)}~.
\label{w6}
\end{eqnarray}

\subsection { Inputs of the model}

The above formalism is applied to study the semi-direct proton decay
of the nuclei $^{41}$Sc, $^{59}$Cu and $^{91}$Nb with the aim of comparing
the predictions with existing data from exclusive measurements.

Although the QPM model Soloviev et al.\cite{Sol_92,V_V_S_S_85}
is not fully consistent since the residual
interaction between quasiparticles is not derived from the
quasiparticle mean field, it has the advantage that its two-body
residual interaction is chosen of a multipole-multipole separable form
and therefore, it
enables one to work with configuration spaces of large dimensions
without facing the cumbersome problem of diagonalizing large matrices.
This is very important for the results to be meaningful as it was already
shown in a previous work on neutron emission\cite{G_S_V_F_96}.

The mean potentials entering the single-particle hamiltonians $h_i$ are
of Woods-Saxon form.
Their parameters are chosen according to Refs.\cite{CHEP_67,TAK_69}
with some readjustments such that,
once the coupling to
phonons is included, the ground states  are
located at their respective experimental positions with respect to the
proton separation threshold. The QPM hamiltonian includes the
monopole pairing which must be taken into account for nuclei with open
shells\cite{G_S_V_88}. This is the case when we calculate the core
properties of $^{59}$Cu and $^{91}$Nb.

The residual particle-hole interaction of Eq. (\ref{f2b}) is
taken of a separable form in coordinate space with effective
interaction strengths considered as adjustable parameters. For the
radial interaction form factor we have used $f(r)=dU/dr$ where $U(r)$
is the central part of the Woods-Saxon potential. For each excitation
mode of the even-even core corresponding to a given angular momentum,
parity and isospin, the interaction strength is found by requiring
that the lowest collective state calculated in RPA be at the
experimental energy. These interaction strengths are also used for the
quasiparticle-phonon coupling.
We have modified the single-particle proton spectrum for $^{91}$Nb in comparison
with Ref.\cite{G_S_V_F_96} to get a better description of
high-lying $1g_{7/2}$, $1h_{11/2}$ subshells, but so far as energies of levels
near the Fermi surface are practically the same there are no changes
in the properties of low-lying vibrational states in comparison with those of
Ref.\cite{G_S_V_F_96}.
For example, the calculated B(E$\lambda$) values of the first
low-lying collective states $3^{-}_{1}$, $5^{-}_{1}$ in
$^{40}$Ca are B(E3)=1.36$e^2fm^6$, B(E5)=2.67$e^2fm^{10}$
that can be compared with the corresponding experimental values 1.24$e^2fm^6$ and
2.97$e^2fm^{10}$. For all nuclei under consideration the
calculated values of B(E$\lambda$) are in general agreement with
experiment. In the present model, these low-lying phonons of the core
are the physical channels where the initial state in the excited odd-A
nucleus can decay by semi-direct proton emission. In the actual
calculation, a very large number of RPA phonons are included in the
quasiparticle-phonon basis, but for $^{40}$Ca the first quadrupole state
cannot be treated as a one-phonon or particle-hole state and it is outside
the present model. On the other hand, the experimental data
\cite{Guil_90} show that the proton decay of $^{41}$Sc
to the $2^+_1$ state in $^{40}$Ca is very weak.

At high excitation energies, when the level density becomes large,
it is often more convenient to calculate the strength function \cite{Boh69}
instead of solving the secular equations (\ref{f10}).
One defines the strength function
$\bar C^2_\alpha(E)$ as the strength distribution
$\vert C^{(i)}_\alpha \vert^2$ folded with an averaging function $\rho$ :
\begin{eqnarray}
\bar C^2_\alpha(E) & = & \sum_i \vert C^{(i)}_\alpha \vert^2 \rho (E
- \Omega_i)~,
\label{m3a}
\end{eqnarray}
where $\rho(z)$ is usually chosen as a Breit-Wigner function, $\rho(z) =
\frac {\Delta}{2\pi} \frac{1}{z^2 + \Delta^2/4}$.
It is shown in Ref.\cite{G_S_V_F_96} that  the calculation of the
strength function $\bar C^2_\alpha(E)$ can be done without the detailed
knowledge of the amplitudes $\vert C_\alpha^{(i)}\vert^2$.
The same averaging procedure can be applied to calculate the partial
escape widths and the branching ratios. In all calculations performed
in this work we have adopted the value $\Delta$ = 0.1 MeV.

\section{Results and discussion}

Using the methods presented above
we have calculated the cross sections and
branching ratios for the non-statistical proton decay of high-lying
states with high angular momenta in $^{41}$Sc, $^{59}$Cu and $^{91}$Nb
going to the ground and low-lying collective states of $^{40}$Ca, $^{58}$Ni
and $^{90}$Zr, respectively.

An example of cross section calculations by the strength function
method is shown in Fig. 1 for the case of proton decay of
the $g_{9/2}$ states in $^{41}$Sc. Since the calculated cross
section is proportional to the square of the one quasiparticle amplitude
of the wave function this figure enables one to see the $1g_{9/2}$ strength
distribution in $^{41}$Sc, too. The $1g_{9/2}$ strength is distributed over
a broad energy interval due to the coupling with the collective vibrations.
Results of our calculations for the $1g_{9/2}$ proton particle strength and
experimental data obtained from the $^{40}$Ca($^{3}$He,dp) reaction at
$E_{^{3}He}$=240 MeV\cite{Guil_90} are presented in Table 1. It is seen
from this table that the calculations reproduce reasonably well the integral
characteristics of the $1g_{9/2}$ proton strength distribution in $^{41}$Sc.

The partial contributions of the non-statistical proton decay
into the ground state and
$3^{-}_{1}$, $5^{-}_{1}$ states of $^{40}$Ca are presented in Fig. 2.
It is seen from Fig. 2
that the partial cross sections are very energy dependent and as a result
different channels can dominate at some excitation energies.
The solid curve in Fig. 1 presents the sum of
all 3 partial channels.
As one can see from  Fig. 2 at excitation energies below 8.4 MeV
the ground state channel dominates, but there are strong transitions
to the $5^{-}_{1}$ state at 7.5 and 8.0 MeV. The $3^{-}_{1}$ and $5^{-}_{1}$
channels begin to contribute importantly starting from 8.4 MeV and they become
the main contributors after 9.5 MeV.

The sum of partial widths for different channels in  some energy intervals
are given in Table 2. As one can see from Table 2 for the energy interval
2.0 - 12.4 MeV the proton decay to the $5^{-}_{1}$ state gives a contribution of
about 54\% and the $3^{-}_{1}$ channel contributes about 37\% in the total sum.
There is a predominance of the $5^{-}_{1}$ channel in comparison with
the $3^{-}_{1}$ one in the energy interval 7.0 - 12.4 MeV.
Such a behaviour can be understood from the structure of the decaying
states and the angular momentum and energy carried away by the emitted
proton.
It is seen from Eq. (\ref{w3}) that the partial escape amplitude
for fixed values of $lj$ is proportional to the contribution of the
quasiparticle-plus-phonon configuration made of a quasiparticle with
the same $lj$ and a phonon of the final state. The cross section depends
on the square of these amplitudes summed over all $lj$ allowed by the
angular momentum coupling rules.
In the case of the proton decay of the $g_{9/2}$ states
at excitation energies 7--12 MeV into the  $3^{-}_{1}$ channel
the outgoing protons carry $(l,j)$=(1,3/2) mainly whereas it
is $(l,j)$=(1,3/2) and $(l,j)$=(1,1/2) for the
$5^{-}$ channel. Thus, penetration factor arguments favour the latter
channel.
It is well known that the strongest coupling between the single-particle
states and phonons takes place for the collective phonons. That is why
the coupling with the low-lying vibrations is mainly responsible of the
damping of the high-lying single-particle modes and giant resonances
in nuclei. This is the case for the $3^{-}_{1}$ and $5^{-}_{1}$ states in
$^{40}$Ca. By this reason the strengths of the $3^{-}_{1}$ and $5^{-}_{1}$
channels are redistributed in a broad energy interval. The decrease of the
spectroscopic strength with an increase of the excitation energy leads to
a decreasing of cross sections of the channels discussed above.
It is worth to mention that a very similar behaviour for
the $3^{-}_{1}$ and $5^{-}_{1}$ channels for the neutron
decay of the high angular momentum states in $^{209}$Pb also takes place
\cite{G_S_V_F_96}.
Our conclusions about the role of different channels for the proton
decay in $^{41}$Sc agree with observations made in Ref.\cite{Guil_90},
but no separation of the direct part from the statistical one for the
proton decay was done in that work.

As a second example we consider the proton decay of the $g_{9/2}$ states
in $^{59}$Cu.
 Results of our calculations for the $g_{9/2}$
strength distribution in $^{59}$Cu and experimental data\cite{NDS_93}
are presented in Table 3. The calculations reproduce correctly the strength
near 7 MeV and this part is mainly responsible for the proton decay.
For the excitation energy interval (2 - 8.9) MeV in $^{59}$Cu the summed
ground state width is equal to 10.3 eV and this is much less than in
the $^{41}$Sc case. An additional contribution from the 2$^+$ channel has
a summed width that is six times less than that of the ground state channel.
There are no transitions to other vibrational states. This is easily
understood if one looks at the structure of the states under discussion.
For the $2^{+}$ channel the outgoing protons carry $(l,j)$=(4,9/2)
and there is a suppression of such transitions due to penetration
factor effects. The configurations including $3^{-}_{1}$ and $5^{-}_{1}$ states
coupled with continuum are located at somewhat higher energies. This is why
no proton decay can proceed to the $3^{-}_{1}$ and $5^{-}_{1}$ states
in $^{59}$Ni.


Very recently, detailed experimental information about proton
decay of isobaric analog states (IAS)
in $^{91}$Nb has been shown in Ref. \cite{Molen}. The IAS
are strongly excited by means of the $^{90}$Zr$(\alpha,t)$$^{91}$Nb reaction.
A sharp peak is seen at 12 MeV excitation energy. Around this excitation energy
three IAS have been
observed earlier\cite{Fink_79} at 11.93 MeV, 12.07 MeV and 12.15 MeV.
It is shown in Ref.\cite{Molen} that at 12 MeV by means of $(\alpha,t)$ reaction
predominantly the $h_{11/2}$ state is excited. The proton decays to the ground
state and some low-lying states are observed. The partial differential
cross sections
for each final state are given. The data reveal proton decays predominantly
to $5^-_1$ and $3^-_1$ excited states of $^{90}$Zr.

Experimental data\cite{Fink_79} for energies and spectroscopic factors of IAS
and results of our calculations are given in Table 4. Besides, this
table also contains calculated partial widths for the proton decay of three IAS
to the ground, 2$^+_1$, 3$^-_1$, 4$^+_1$ and 5$^-_1$ states.

As one can see from Table 4 the calculated excitation energies for the $7/2^+$
IAS are higher than the experimental ones but the spectroscopic factors
extracted from experimental data within 20\% accuracy are reproduced
reasonably well.

Let us discuss the proton partial widths for these IAS. It is seen from
Table 4 that for the first $7/2^+$ state the main decay channel is the
$4^+_1$ channel. For this state the $(l,j)$=(0,1/2) protons are most
important for the $4^+_1$ channel and there is a rather weak transition
for the $2^+_1$ channel due to the $(l,j)$=(2,3/2) protons.
The ground state width is proportional to the one quasiparticle strength
(the spectroscopic factor). Since the contribution of the one
quasiparticle component in the norm of the wave function of this IAS is
much less than contributions of the quasiparticle-plus-phonon components
the transition to the ground state is weaker than to the $4^+_1$.
The particularities of the partial proton widths for the second
$7/2^+$ state can be understood from the structure of this state, too.
For the $4^+_1$ channel the $(l,j)$=(2,3/2),(2,5/2) and (4,7/2) protons can
contribute besides the $(l,j)$=(0,1/2) protons. As a result the decay
width to the $4^+_1$ state increases strongly for the second $7/2^+$
IAS in comparison with that of the first $7/2^+$ state.
An increase of the contribution of the configuration constructed from
the first quadrupole phonon and $2d_{3/2}$ single-particle state
in comparison with their contribution to the structure of the first
$7/2^+$ state results in an essential growth of the decay width to
the $2^+_1$ channel. The ground state width becomes larger mainly because of
the increased spectroscopic factor. The transitions
to the $5^-_1$ state can take place due to the outgoing protons with
$(l,j)$=(5,11/2) but they are supressed because of penetration
factors and a small contribution of relevant components in the
wave function structure.

In the case of the $11/2^{-}$ IAS the $5^-$ channel dominates and this
is due to the outgoing protons carrying $(l,j)$=(0,1/2),(2,3/2) and
(2,5/2). The components containing the $3^-_1$ phonon have a small
contribution in the wave function structure and as a result the
width for the $3^-_1$ channel is small, too.  In spite of unfavourable
penetration factor the $2^+_1$ width is 4 times larger than that of the $3^-_1$
channel. This is due to the configuration consisting of the $2^-_1$
phonon coupled with $h_{11/2}$ which contributes about 15\% in the norm
of the $11/2^{-}$ IAS.

The comparison of the calculated partial widths with the data
of Ref.\cite{Molen} shows that there is a qualitative agreement.
The main channels of the proton decay of (11/2)$^-$ IAS are reproduced
in the calculated structure. The dominance of the $5^-_1$ channel is
well established and the calculated partial width of 3.3 keV (Table 4)
is in agreement with the measured one (2.9 keV).
The $3^-_1$ channel is more pronounced than ground state channel but the
calculated partial widths for both channels are much less than
the measured ones. The calculations indicate a large decay to
the $2^+_1$ channel.
Such decay is discussed in Ref.\cite{Molen} but quantitative estimations
for the contribution of the $2^+_1$ channel in the cross section
have not been evaluated in that work.

The summed partial widths for the proton decay of the three IAS to the
low-lying vibrational states are also presented in Table 4. As one can see
from the last row of Table 4 the $4^+_1$ channel exhausts almost 60\%
of the total cross section, the $5^-_1$ channel contributes about 26\%
and the $2^+_1$ channel gives about 10\% of the total strength. The
contribution of the ground state channel for the proton decay of the three
IAS in $^{91}$Nb is 4\% only.

\section{Conclusions}

A microscopic approach based on the QPM has been applied to calculate
the non-statistical proton decays of high angular momentum states excited in
one-nucleon transfer reactions.
Partial cross sections and branching ratios for proton emission from
high-lying states in $^{41}$Sc, $^{59}$Ni and $^{91}$Nb
 have been evaluated.
The calculated branching ratios demonstrate the existence of
strong direct proton decays to the low-lying vibrational states
in $^{41}$Sc and $^{91}$Nb and enables one to understand particularities
of the decay to different channels.
One can conclude from an analysis of calculated partial cross sections
that, for high angular momentum states the non-statistical
proton decay is more favourable
into the higher angular momentum and lower excitation energy final states
when the two following conditions are fulfilled: a strong
particle-phonon coupling in that channel and a penetration factor
which is not hindered by angular momentum or energy. A similar conclusion
was also reached for the neutron decay case\cite{G_S_V_F_96}.
A general agreement with existing experimental data is obtained and the
predicted branching ratios for different channels can be used to
analyze future experimental data.

\acknowledgements

We would like to thank  G. Crawley and S. Fortier for fruitful discussions
and correspondence.
Ch.S. and V.V.V. thank the hospitality of IPN-Orsay where the main part
of this work was done. This work is partly supported by IN2P3-JINR agreement
and by the Bulgarian Science Foundation (contract No Ph. 801).

\begin{figure}
\caption{Cross sections (arbitrary units) for the excitation and
proton decay
of $g_{9/2}$ states in $^{41}$Sc into low-lying states, as a function
of excitation energy $E$. The solid line is the sum of ground state,
$3^-_1$ and $5^-_1$  channels, the dotted line shows
the partial contribution of the ground state channel.}
\end{figure}

\begin{figure}
\caption[]{Partial cross sections (arbitrary units) for the excitation and
proton decay
of $g_{9/2}$ states in $^{41}$Sc as a function
of excitation energy $E$.
The solid line is the $3^-_1$ channel, the dotted-dashed line is
$5^-_1$ channel and the dotted line is the ground state channel.}
\end{figure}

\begin{table}[]
\caption[]{The $g_{9/2}$ proton particle strength in $^{41}$Sc
.}
\begin{center}
\begin{tabular}{ccccc}
$\Delta E_x$(MeV) &$\overline{E_x}(exp.)$ (MeV) &
$\overline{E_x}(calc.)$ (MeV) &$C^{2}S(exp.)$&$C^{2}S(calc.)$
\\
\hline
2.0 - 12.4  &8.80   &8.96   &0.84 & 0.92  \\
8.4 - 12.4  &10.4  &9.49  &0.56 & 0.76 \\
            &5.04   &4.65  &0.12 & 0.05 \\
\end{tabular}
\end{center}
\end{table}

\begin{table}[]
\caption[]{Summed partial widths (in MeV) for different channels
for the proton decay in $^{41}$Sc}
\begin{center}
\begin{tabular}{cccc}
$\Delta E_x$(MeV) & $\sum\Gamma_{g.s.}^{\uparrow}$ &
$\sum\Gamma^{\uparrow}_{5^{-}_{1}}$&
$\sum\Gamma^{\uparrow}_{3^{-}_{1}}$
\\ \hline
2.0 - 7.2 &0.225 &3.63 $\over$ 10$^{-5}$ & 2.14 $\over$ 10$^{-2}$  \\
2.0 - 8.4 &3.314 &0.170 &2.31 $\over$ 10$^{-2}$  \\
8.4 - 12.4 &1.689 &11.36 &7.832  \\
2.0 - 12.4 &5.003& 11.53& 7.855\\
\end{tabular}
\end{center}
\end{table}

\begin{table}[]
\caption[]{The $1g_{9/2}$ proton particle strength in $^{59}$Cu
.}
\begin{center}
\begin{tabular}{ccc}
$E_x$(MeV) & $C^{2}S(exp.)$ & $C^{2}S(calc.)$ \\
\hline
3.0  & 0.35   & 0.60   \\
$\sim$7.3  & 0.35  &0.35  \\
3.1 - 7.2   &0.09  &0.05 \\
\end{tabular}
\end{center}
\end{table}

\begin{table}[]
\caption[]{Partial widths (in keV) for different channels
for the proton decay in $^{91}$Nb}
\begin{center}
\begin{tabular}{cccccccccc}
$E_x(exp.)$(MeV) & $E_x(calc.)$(MeV)& J$^{\pi}$&
$C^{2}S(exp.)$ & $C^{2}S(calc.)$&
$\Gamma_{g.s.}^{\uparrow}$ &$\Gamma_{2^{+}_{1}}^{\uparrow}$ &
$\Gamma_{3^{-}_{1}}^{\uparrow}$ &$\Gamma^{\uparrow}_{4^{+}_{1}}$&
$\sum\Gamma^{\uparrow}_{5^{-}_{1}}$
\\ \hline
11.79&12.00 &7/2$^+$& 0.11 & 0.18 &0.156& 0.019& 0&0.990&0.005 \\
12.07 &12.15 &11/2$^-$ &0.35& 0.57& 0.049& 0.296&0.073& 0.002&3.308 \\
12.15 &12.90 &7/2$^+$ &0.40&0.29&0.309&1.010&0&6.738&0.027  \\
\hline
& & & &$\sum\Gamma^{\uparrow}$ &0.514&1.325&0.073&7.73&3.340\\
& & & &\% of total $\sum\Gamma^{\uparrow}$&4.0&10.2&0.6&59.5&25.7
\end{tabular}
\end{center}
\end{table}

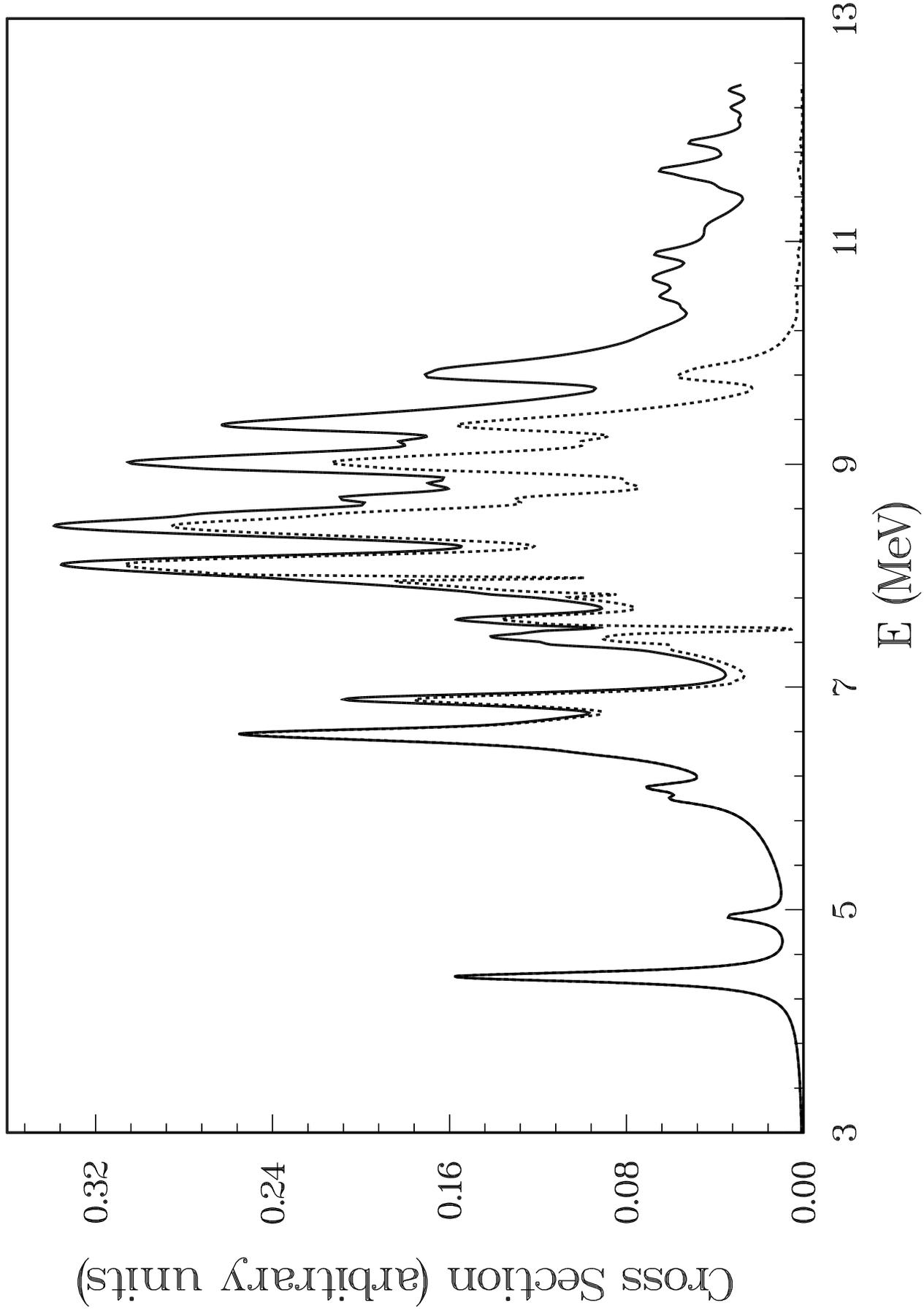
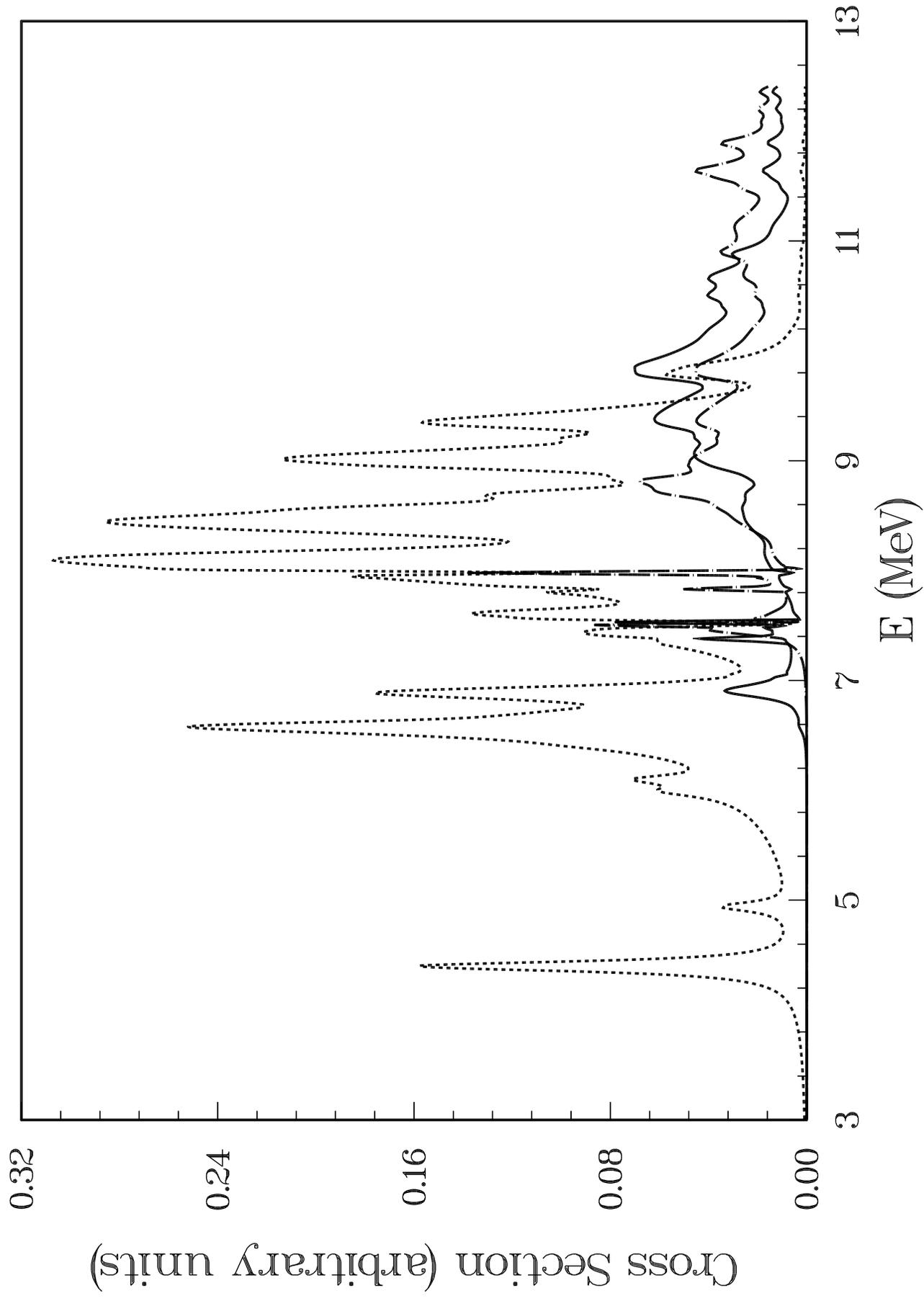
\end{document}